\documentclass[11pt,a4paper]{article}
 \usepackage{graphicx}
 \usepackage{afterpage}
 \usepackage{enumerate}

\begin{document}
\small
 \title{\bf Macro-microturbulence in the solar photosphere
}
\author{V.A. Sheminova }
  \date{}

 \maketitle
 \thanks{}
\begin{center}
{Main Astronomical Observatory, National Academy of Sciences of Ukraine,\\ Akademika  Zabolotnoho 27,  Kyiv,  03143, Ukraine,  e-mail: shem@mao.kiev.ua
}
\end{center}

\begin{abstract} {The velocity distribution of the large and small-scale  motion in solar photosphere has been obtained by crossing method based on fitting the observed and calculated equivalent widths as well as the central depths of the spectral lines at the center of the and the limb of the solar disk. We used about 200 Fe I lines. According to our results  the motions in  photosphere are anisotropic. The radial component of microturbulent velocity  decreases from 1.0 to 0.3 km/s and    the tangential one  from 1.7 to 1.3 km/c at the photosphere heights from 200 to 500 km ($\log \tau_5=-1.4$ and $-3.5$).  At the same heights  the radial component of  the macroturbulent velocity decreases from 1.8 to 1.2 km/s and the tangential one from 2.3 to 0.8 km/s.
}
\end{abstract}

{\bf Keywords:} {Sun, spectral lines, turbulence, velocity fields, photosphere
\vspace {1cm}

\section{Introduction}

The macro-microturbulent model of the velocity field in the solar photosphere found wide application in calculating the profiles of spectral lines. As to the photospheric microturbulence, quite a few number of studies have been already published, while the macroturbulence was determined little. Distribution of macroturbulent velocities with height was studied  by Kondrashova [5] and Kostik [6]. They applied the same method using central  depths of the Fe I lines. The recently Sheminova [7] have developed new crossing method.  It is based on the well-known peculiarity of the spectral line, namely: the increase of the equivalent width and decrease of the central depths with increasing microturbulent velocity. In the  previous paper  [9] we determined only the microturbulence $V_{mic}$ by the crossing method, given the macroturbulence obtained by Kostik [6]. The aim of this paper is to derive micro- and macroturbulent velocities in the solar photosphere, using as many Fe I lines as possible and specifying as accurately as possible the iron abundance.

\section{Method and initial data}

The crossing method can be explained using Fig. 1. The line parallel to abscissa axis corresponds to the adopted iron abundance $A= \log N_{Fe} /N_H + 12= 7.64$.  The $A_W$ line and the $A_d$ line  corresponds the abundance obtained from the observed equivalent width $W$ and central depth $d$ of a spectral line using a series of $V_{mic}$ values. A set of $A_{d}$ lines is calculated for the series $V_{mac}$ until the $A(7.64)$ line, the $A_W$ line and one of the $A_{d}$ lines intersect at one point.   The crossing point determines $V_{mic}$ and  $V_{mac}$.  Fits for different spectral lines formed at  different photospheric heights give the height dependencies of $V_{mic}(h)$ and $V_{mac}(h)$. Disadvantage of the method: the abundance $A$ and the oscillator strengths $\log gf$ used should be known precisely.

In our calculation was used  the HOLMU photospheric model of Holweger and Muller [4], $\gamma = 1.5 \gamma_{\rm WdW}$, the oscillator strengths of Gurtovenko and Kostik [2]. The iron abundance of 7.64 was previously obtained by us in [8] from the equivalent widths of  weak lines observed in the centre of the solar disk ($\cos \theta = 0$) and  on the limb  ($\cos \theta = 0$). The height of line formation was  calculated using the depression contribution functions detail described by Gurtovenko and Sheminova [3]. The formation height of  whole line profile $h_W$ is determined by the weighted mean over all the heights of formation  at the specific depression $d$ within the line profile. The height corresponding to $V_{mic}$ and  $V_{mac}$ velocities was determined as the average  of the formation height of the whole line profile $h_W$ and the formation  height of its central depth $h_d$.

The equivalent widths and central depths of Fe I lines were take from original center-to-limb observations at Kiev Observatory with the double-pass spectrometer. The number of all spectral Fe I lines was about 200 with the central depths $d \gg 20$\%. All lines were divided into groups with a small range in height  of formation.

Within the accepted approximations the photospheric model, damping constant, the adopted $A$ and $\log gf$ may introduce some uncertainty into the results. Inaccuracies in the damping constants appreciably influence the results obtained from lines of high excitation potential. The corresponding maximum error of the velocities  can reach 0.3--0.4 km/s. Different solar photospheric models give appreciable differences in the velocities about 0.4 km/s. Parameters  $A$ and $\log gf$ act as a corresponding shift of the straight line $A =$~const.
 \begin{figure}[!tb]
 \centerline{
\includegraphics [scale=0.7]{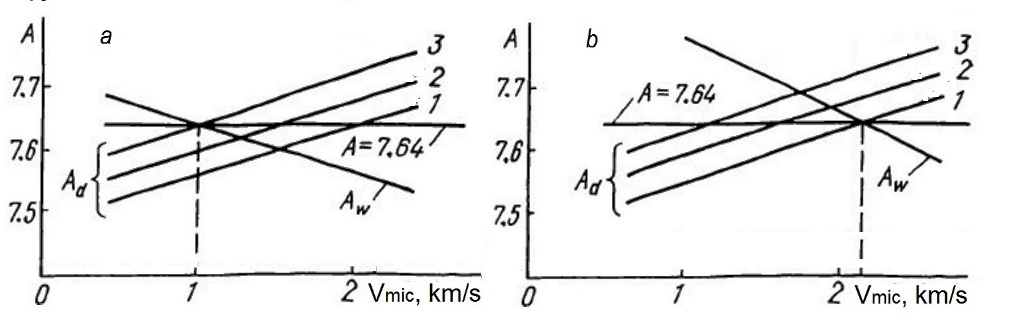}
     }
  \caption {\small
 An example of the determination of $V_{mic} $ and $V_{mac} $ by a crossing method using the Fe~I~6290~{\AA} line in the disc center (a) and at the limb (b). $A_{W}$  line corresponds the iron abundance derived from the equivalent widths for different values of $V_{mic} $. $A_{d}$ lines correspond the abundance derived from the central depths for different values of $V_{mic} $ and $V_{mac} = 1.09$ (1), 1.39(2), 1.9 km/s (3). The line parallel to abscissa axis corresponds to the adopted iron abundance $A = 7.64$. The point of intersection of  $A_{W}$, $A_{d}$, and $A = 7.64$ lines determines the $V_{mic}$ and $V_{mac} $ values.
}
\end{figure}

\section{ Results and discussion}

For each group of lines we have obtained the average values of $V_{mic} $ and $V_{mac} $  and its r.m.s. error. All the results are given in Figs. 2 and 3 for $\cos \theta = 1$ (radial component) and $\cos \theta = 0.3$ (tangential component). The main features of the micro-macroturbulence in the solar photosphere  are as follows.

{\bf Microturbulence.}  The amplitude of the radial component decreases with height from 1.2 to 0.3 km/s at heights of 150 ($\log \tau_5 = -1.4$) and 450 km ($\log \tau_5 = -3.5$).  The amplitude of the tangential component is greater than the radial component. It decreases with a height from 1.7 to 1.2 km/s respectively at heights of 200 and 600 km.  The mean square error is $\pm 0.3$ km/s.

{\bf Macroturbulence.}  The amplitude of the radial component decreases with a height from 2.5 to 1.2 km/s at heights of 150--500 km.  The amplitude of the tangential component decreases from 2.2 to 0.8 km/s at heights of 200--500 km and 1.0 km/s at  600 km, and it  increases at heights greater than  600 km. The mean square error is $\pm 0.5$ km/s.
\noindent
 \begin{figure}[!h]
 \centerline{
\includegraphics [scale=0.5]{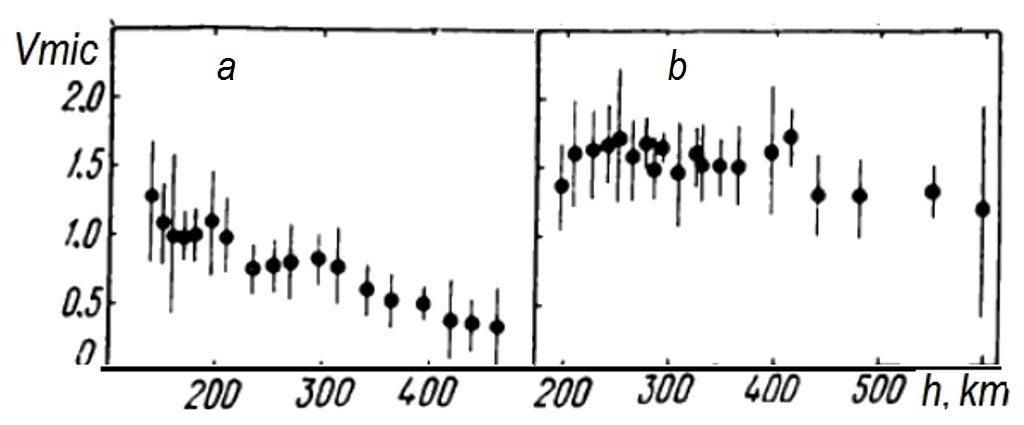}}
      \caption {\small
 Results of determination of microturbulence $V_{mic}$ by the crossing method. The radial component (a) and tangential component (b) depending on geometric height $h$ ($h=0$ km at $\log \tau_5 =0$). Each point ($\bullet$) represents the average microturbulent velocity  and the depth of line formation  within a specific group. Vertical bars, doubled r.m.s. errors.}
\end{figure}
 \begin{figure}[]
 \centerline{
\includegraphics [scale=0.5]{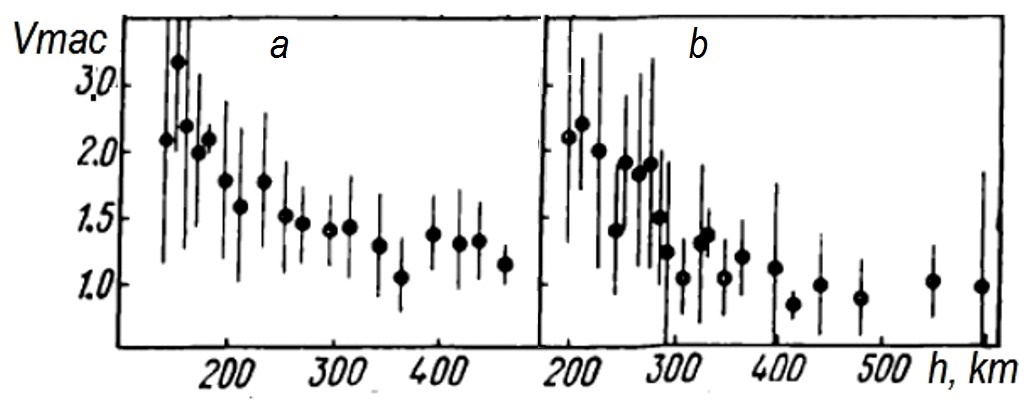}}
    \caption {\small
 The same as Fig.2 but for macroturbulent velocity $V_{mac}$.
 }
\end{figure}

 It should be noted the relatively large errors of the results for deep layers, especially for $V_{mac} $. The growth of errors in the region of formation of weak lines ($h = 200$--300 km) is clearly seen in Fig. 3. This is probably due to the low sensitivity the line depths  to the macroturbulence  as well as the equivalent widths to the microturbulence for  weak lines. Our calculations show that we must  select  the spectral lines with central depths from 50\% to 80\% to increase the accuracy of determining $V_{mac} $ by this method.

Let's turn to Fig. 4 which shows the dependencies $V_{mac} (h)  $ obtained by Kondrashova [5] (curve 1) and  Kostik [6] (curve 2), as well as in this work (curve 3). Curve 1 is plotted according to the results of the analysis of 437 lines, the central depths of which are $\geq 5$\%, and curve 2 is plotted according to 21 lines with $d\geq 35$\%. All other  parameters used in [5] and [6] are the same: HOLMU model, $A = 7.59$, $\gamma = 1.3 \gamma_6$, $V_{mic}^{rad} = 0.8$ km/s, $V_{mic}^{tg} = 1.4$ km/s, and $\log gf$ [2]. However, as can be seen from Fig. 4 the curves 1 and 2  differ each other. The radial components differ  in the region $h\leq 200$ km while the  tangential ones in the region $h> 200$ km. The curve 3 obtained in this paper is closer to curve 2 [6].

As for $V_{mic} (h)$ obtained in this paper simultaneously with $V_{mac} (h)$, it fits quite well into the family of $V_{mic} (h)$ curves obtained by different methods based on the Fraunhofer spectrum.  The average value of the radial component of microturbulent  velocity also agrees well with recent results obtained from equivalent widths (Sheminova and Gurtovenko [10], Blackwell et al. [1], Kostik [6]).

 \begin{figure}[!h]
 \centerline{
\includegraphics [scale=0.42]{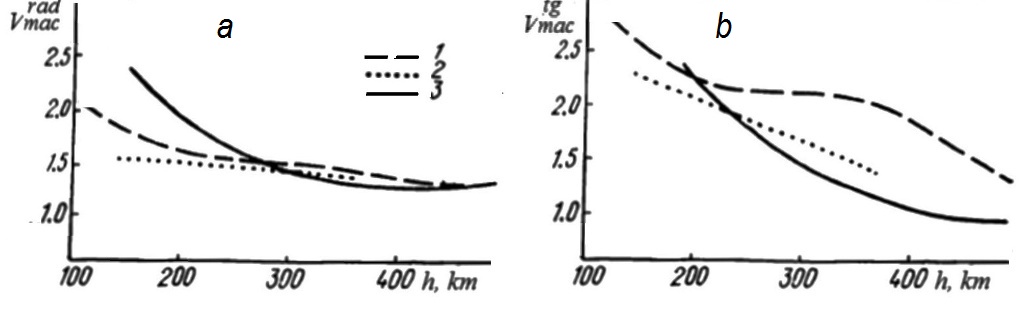}
     }
 \caption {\small
  The radial (a) and tangential (b) components of macroturmulence  according to the data Kondrashova [2] -- 1, Kostyk [3]  -- 2, real research -- 3.
}
\end{figure}


\section{Conclusions}

 The application of a new method based on the use of the observed equivalent widths and central intensities of spectral lines made it possible to simultaneously find the distribution of the velocities of small- and large-scale motions in the photosphere Sun and stars.

We obtained  new dependence of macroturbulent velocity with height in solar photosphere which confirm the data obtained by Kondrashova [5] and Kostyk [6], namely: the character of the motions is anisotropic; the amplitude of the velocities decreases with height in the photosphere; the gradient of the tangential component is greater than the radial one.  The obtained new dependence of the microturbulent velocity with height in the solar photosphere confirms in general terms the dependencies obtained earlier.

The accuracy of determining the macroturbulent velocities  by the crossing method reaches on average $\pm0.5$ km/s and for the microturbulent velocities it is $\pm0.3 $ km/s. The use of weak lines with central depths $d<20$\% leads to relatively large errors of the results for deep layers of the photosphere ($h = 200$--300 km).

 The crossing method may be applied when studying the chemical composition of stars using spectral lines if the equivalent widths, central intensities, oscillator strengths are known  but information about macro- and microturbulent velocities and abundance of chemical elements   is not available. In such cases the calculation schema will be more complicated. First, one derive the iron abundance from the observed equivalent widths of the weak Fe I lines.  Second, one derive $V_{mic} $ and $V_{mac} $ from the moderate and moderately strong Fe I  lines using the crossing method. Thirdly, having  $V_{mic} $ and $V_{mac} $  for a particular star one can obtain  the abundances of any chemical elements by any available method.

\vspace {0.3cm }
The author is grateful to T. V. Orlova for her help in the calculations.
\vspace {0.3cm }

\end{document}